\documentstyle[12pt,fleqn]{article}
\newcommand{\eqnsection}{
\renewcommand{\theequation}{\thesection.\arabic{equation}}
\makeatletter
\csname $addtoreset\endcsname
\makeatother}
\input{dlat}
\begin{document}
\eqnsection
\title{ New integrable hierarchies gauge generated from
constrained KP
  }
\author{Anjan Kundu$^\dagger$  and  Walter Strampp
 \\
Fachbereich 17--Mathematik/ Informatik \\
GH--Universit\"at Kassel \\
Holl\"andische Str. 36 \\ 34109 Kassel, Germany
}
\date{}
\maketitle
\begin{abstract}
Exploiting the residual gauge freedom in the formulation of constrained
 KP hierarchy a number of new integrable
systems are derived including
hierarchies of   Kundu-Eckhaus equation and
 higher order nonlinear extensions of
Yajima-Oikawa  and  Melnikov hierarchy. In the multicomponent
case such
 gauge freedom generates novel
 multicomponent   as well as vector generalisations of the above
 systems, while the constrained modified KP hierarchy  is
found   to yield another set of equations like derivative NLS,
Gerdjikov-Ivanov equation and chen-Lee-Liu equation depending
 on the gauge choice.
\end{abstract}
\\ \\ \\
Running title: {\it  Integrable hierarchies from constranied KP
 }
\\ \\ \\ \\ \\ \\ \\  \\
\hrule
{\it $^{\dagger}$Permanent address}:
 Saha Institute of Nuclear Physics, AF/1 Bidhan Nagar, Calcutta 700 064,
  India.

\newpage
\section{Introduction}
\setcounter{equation}{0}
It is well established that the scalar KP hierarchy allows a series
of constraints [1-3] expressed through its symmetry  $
                        (\sum_{i=1}^m \Psi^i \Psi^{i*})_x$, where
      $\quad\Psi^i$ and $ \ \Psi^i, \ i = 1,\cdots m
\quad$ are the eigenfunctions and
adjoint eigenfunctions of the corresponding linear system.
In particular, the constraints of the type
\begin {equation}
u_x=  (\sum_{i=1}^m \Psi^i \Psi^{i*})_x , \ \
u_y=   (\sum_{i=1}^m \Psi^i \Psi^{i*})_x
  \ \ \mbox{and} \ \ u_t= (\sum_{i=1}^m \Psi^i \Psi^{i*})_x
,\end {equation}
are known to produce multicomponent
  AKNS , Yaijima-Oikawa and Melnikov hierarchies,
 respectively [3]. A simple but crucial observation, which inspired the
 present investigation is that the constraint (1.1) does not
 fix uniquely the
   resultant equations obtained for $ \Psi^i$ and $ \Psi^{i*}$
 and allows a residual gauge freedom
   $ \tilde \Psi^i=e^{-\theta} \Psi^{i} ,\ \tilde \Psi^{i*}= \Psi^{i*}
 e^{\theta} $ with $\theta$ being an arbitrary function
 , which can be exploited
 further to generate  gauge transformed integrable systems. Moreover,
 in the multicomponent case one can have different gauge functions
 $\theta^i$ for different components allowing a much wider choice.
Our aim here is to exploit such gauge freedom to obtain  new
integrable  hierarchies like Kundu-Eckhaus [4-6] hierarchy, higher-order
Yaijima-Oikawa as well as higer-order Melnikov hierarchy.
Corresponding  multicomponent cases would yield  novel  multicomponent
as well as vector
extensions of the above systems. It is worthwhile to note that
multicomponent  extensions exhibit an interesting breaking of
global symmetry , while still  preserving the integrability of the system.
Similar gauge freedom is found to be valid equally in the case of
 constrained modified KP hierarchy, investigated in the recent past [7,8]
 and moreover
the possibilities  of  gauge generation
 from such a system found to be even richer.
 The   simplest constraint   $v= \Psi \Psi^* $ in this case can yield
a series of different equations like  derivative NLS  [9],
 Gerdjikov-Ivanov [10] and Chen-Lee-Liu [11] equations depending
 on particular  gauge choice.

The organisation of the paper is as follows. In sec.2 we describe the
  main idea of the gauge transformation in the simplest case  and
  derive the hierarchy   of Kundu-Ekchaus equation.
Sec 3. gives a systematic  way for such gauge choice in a
 general framework and obtains
 hierarchies of   higher-order  extended
Yaijima-Oikawa and Melnikov equations.
 In sec 4. we construct multicomponent and vector generalisation of the above
systems
, while sec. 5 presents such application to the constrained modified KP
hierarchy.
 Sec. 6  gives  concluding remarks.

\section{ Constrained KP and generation of Kundu-Eckhaus hierarchy}
\setcounter{equation}{0}
 It is known [12] that the KP hierarchy can also be formulated through
 Zakharov-Shabat type linear system
\begin{equation}
  \Psi_{t_l}= B_l \Psi,  \ \  \Psi^*_{t_l}= -B_l^* \Psi^* ,\ \ l=1,2,\cdots
  \end{equation}
with   infinite series of $B_l$ operators given by
\begin{eqnarray}
B_1 & = & \partial_x  \,,\\
B_2 & = & \partial_x^2 + 2u_2  \,,\\
B_3 & = & \partial_x^3 + 3u_2\partial_x + 3u_3 + 3u_{2,x} \,, \\
B_4 & = & \partial_x^4 + 4u_2\partial^2_x + (4u_3 + 6u_{2,x})\partial_x
          + 4u_4 + 6u_{3,x} + 4u_{2,xx} + 6u_2^2\,.
\end{eqnarray}
etc., the compatility condition  of (2.1)
 \begin{equation}
   B_{l,t_k}-  B_{k,t_l} +
  [B_{l}, B_{k}] =0
 \end{equation}
 yielding subsequently the
 corresponding field equations along with differential
  relations among various
 fields $u_i$  as
\begin{eqnarray}
u_3 & = & -\frac{1}{2}u_x+\frac{1}{2}\partial^{-1}_xu_y\,,\\
u_4 & = & \frac{1}{4}u_{xx}-\frac{1}{2}u_y -
          \frac{1}{2}u^2+\frac{1}{4}\partial_x^{-2}u_{yy}\,.
\end{eqnarray}
etc.
  As a result all equations of this hierarchy can be
 expressed  finally through a single variable $u_2\equiv u$ only.
It has also been observed
 that the product $\Psi\Psi^*$ acts as a
generator of conserved densities and symmetries of the KP hierachy,
which permits to impose constraints on this system by expressing
the dynamical field $u$ through    $\Psi\Psi^*$ [1-3].
Let us consider the simplest of such constraits given by
        \begin{equation}
 u= (\Psi\Psi^*)
 ,\end{equation}
which  yields from (2.1-5), as is seen readily, the AKNS type
 equation
for $l=2$ and its hierarchy for $l>2$.
It is important to observe now that the freedom in this constraint is not
fully exhausted in the system (2.1) with operators (2.2-5) and the
residual local gauge  can be used by defining a transformed set of
eigenfunctions   as $\tilde \Psi= h^{-1} \Psi,  \ \ \tilde
 \Psi^*= \Psi^* h $ ,
  where $ h $ can be any arbitrary
  function. It should be noted however that
 under such a change the field $u$
 defining the constraint (2.9)
  remains the same solution of the  KP hierarchy,
 while the corresponding AKNS hierarchy derived as the contrained system
gets  gauge transformed into
\begin{equation}
\tilde  \Psi_{t_l}= \tilde B_l \tilde \Psi,  \
 \  \tilde \Psi^*_{t_l}= - \tilde B_l^*  \tilde \Psi^* ,\ \ l=1,2,\cdots
  \end{equation}
       with the transformed operators
       \[\tilde B_l= h^{-1}B_l h -h^{ -1} h    \]
 along with the constraint (2.9). Our other crucial
     observation suggested by the operator contents of  $B_l$ is that,
     the  gauge function chosen in the form $h=  e^{\theta} $  simplifies
     the  structure of the transformed operators $\tilde B_l$ considerably
    by cancelling the function $ e^{\pm \theta}$
      from  all the expressions of $\tilde B_l $ and
  thus   one  expects to obtain some physically interesting systems in
  this gauge.
     For example,  we get now
      for $t_2\equiv y$ the form
        \[ \tilde B_2= B_2 -{\theta}_y +{\theta}_x^2
     +2{\theta}_x \partial+{\theta}_{xx}\] giving the  constrained system
     as
\begin{equation}
\tilde  \Psi_{y}= \tilde  \Psi_{xx}+2 u
 \tilde  \Psi +(
 -{\theta}_y +{\theta}_x^2  +{\theta}_{xx} )\tilde  \Psi
     +2{\theta}_x
\tilde  \Psi_x
,\end{equation}
\begin{equation}
\tilde  \Psi_{y}^*= -\tilde  \Psi_{xx}^* -2 u
 \tilde  \Psi^* +(
 {\theta}_y -{\theta}_x^2  +{\theta}_{xx} )\tilde  \Psi^*
     +2{\theta}_x
\tilde  \Psi_x^*
,\end{equation}
and similarly for higher evolutions:
\begin{eqnarray}
\tilde{\Psi}_t& = &\tilde{\Psi}_{xxx} + 3u\tilde{\Psi}_x +
                \frac{3}{2}u_x \tilde{\Psi}+
                \frac{3}{2}\left(\partial_x^{-1}u_y\right)
                \tilde{\Psi} \nonumber \\&+&
                 (-\theta_{t} +3u\theta_{x} +\theta_{xxx}
                 +3\theta_{x}\theta_{xx}+\theta_{x}^3)\tilde \Psi \nonumber
                 \\ &+& 3 (\theta_{xx}
                  +\theta_{x}^2)\tilde \Psi_x+3\theta_{x}
                 \tilde \Psi_{xx} \ ,
                  \end{eqnarray}
                   \begin{eqnarray}
\tilde{\Psi}^*_t& = &\tilde {\Psi}^*_{xxx} + 3u\tilde {\Psi}^*_x +
                  \frac{3}{2}u_x \tilde {\Psi}^*-
                  \frac{3}{2} \left(\partial_x^{-1}u_y\right)
                  \tilde {\Psi^*}
                  \nonumber \\&-&
                 (-\theta_{t} +3u\theta_{x} +\theta_{xxx}
                 -3\theta_{x} \theta_{xx}+\theta_{x}^3)\tilde \Psi^* \nonumber
                 \\ &-&3(\theta_{xx} -\theta_{x}^2)
                  \tilde \Psi_x^*-3 \theta_{x}
                 \tilde \Psi_{xx}^*
\end{eqnarray}
 etc..
It may be observed further that for the particular gauge choice (with the
notation $t_1\equiv x, t_2 \equiv y , t_3\equiv t $)
       \begin {eqnarray}
{\theta}_x  &=&   \tilde \Psi \tilde \Psi^* =u
 ,  \\
{\theta}_y  &=&    (\tilde \Psi_x\tilde \Psi^*-\tilde \Psi\tilde \Psi^*_x)
+
2 (\tilde \Psi\tilde \Psi^*)^2,
    \\ {\theta}_t &=&
    \tilde \Psi_{xx} \tilde \Psi^*+
   \tilde \Psi\tilde \Psi^*_{xx}  -\tilde \Psi_x \tilde \Psi^*_{x}
+
3 (\tilde \Psi\tilde \Psi^*)^2
\nonumber \\ &+& 3      ( \tilde \Psi\tilde \Psi^* )
  (\tilde \Psi_x\tilde \Psi^*-\tilde \Psi\tilde \Psi^*_x)
+  3
 (\tilde \Psi\tilde \Psi^*)^3
,\end{eqnarray}
consistency of which: $\theta_{t_it_j}=  \theta_{t_jt_i} $ may be checked
directly using the equations  (2.11-14) together with the constraint (2.15),
  we derive
finally the higher-order nonlinear extensions to AKNS hierarchy as
 \begin{equation}
\tilde  \Psi_{y}= \tilde  \Psi_{xx}+\left(2 (\tilde  \Psi\tilde  \Psi^*)
+ 2(\tilde  \Psi\tilde  \Psi^*)_x -(\tilde  \Psi\tilde  \Psi^*)^2
\right) \tilde  \Psi
,\end{equation}
    \begin{eqnarray}
\tilde  \Psi_{t}&=& \tilde  \Psi_{xxx}+ \left(6 (\tilde  \Psi\tilde  \Psi^*)
  + 3 (\tilde  \Psi\tilde  \Psi^*)^2 \right)\tilde  \Psi_x \nonumber \\
&+&\left(3
 (\tilde  \Psi\tilde  \Psi^*)^2 -2(\tilde  \Psi\tilde  \Psi^*)^3
   +6(\tilde  \Psi_x\tilde  \Psi^*_x) \right)\tilde  \Psi \nonumber \\
&+& 3 (\tilde  \Psi\tilde  \Psi^*)  \tilde  \Psi_{xx}
      +   3  \tilde  \Psi^2_x \tilde  \Psi^* +6    (\tilde  \Psi\tilde  \Psi^*)
       \tilde  \Psi^2\tilde  \Psi^*_x  \
,\end{eqnarray}
etc. and similar equations for $\Psi^*$. One immediately recognises (2.18)
as the  Kundu-Eckhaus equation and consequently the related higher
 evolution equations
 (2.19) etc. as the corresponding hierarchy. This  shows that the solution
 $\tilde \Psi, \tilde \Psi^*$ of these equations,
 which are each a $(1+1)$-dimensional system, can construct
  through the constraint
 (2.9)
 a solution of the KP hierarchy.
 This at the same time is also true for the solutions of
 the original AKNS hierarchy.
 It may appear that the  choice of $\theta$ made in (2.15-17)
  is somewhat ad hoc
 in nature. However  we show
  in the next section how such  gauge choices can
 be made consistently in a more general
 constrained system, relating them to the conserved densities.
\section {Consistent gauge choice and generation of
                higher-order
Yaijima-Oikawa and Melnikov hierarchy }
\setcounter{equation}{0}
For finding out physically interesting gauge functions in a consistent
way we show  a simple but important relation between the conserved
densities $\tilde \rho_l$
 of the constrained
  system. We notice first that
   in  one hand conserved densities are connected with
  the time evolutions of $u_2\equiv u$ as
\begin {equation}
u_{t_l}=  \partial_x
\tilde \rho_l ,\quad \ \ l=1,2,\cdots
,\end {equation}
while on the other hand  different constraints of the KP hierarchy
like (1.1)
may be given by
    \begin {equation}
  \partial_{t_k}u =  \partial_{x} (\tilde \Psi\tilde\Psi^*)
\ , \qquad\ k=1,2,3,\cdots.\end {equation}
Recall that the simplest constraint with $k=1$ has been considered
already in the last section , while the other cases will be dealt below.
Comparing (3.1-2) one easily obtains
\[ \partial_{t_k}u_{t_l} =
 \partial_x(  \partial_{t_k}
\tilde \rho_l)=
 \partial_{x} (\tilde \Psi\tilde\Psi^*)_{t_l} \] or
    \begin {eqnarray}
   \partial_{t_k}
\tilde \rho_l&=&
  (\tilde \Psi\tilde\Psi^*)_{t_l} \nonumber \\ &=&
      \tilde \Psi_{t_l}\tilde\Psi^*+ \tilde \Psi\tilde\Psi^*
      _{t_l}
      = \tilde B_{l}\tilde \Psi\tilde\Psi^*-\tilde B^*_{l} \tilde\Psi^*
      \tilde \Psi
\end {eqnarray}
using the linear system (2.10). Differentiating  (3.3) once we obtain
 further
     \begin {eqnarray}
   \partial_{t_k}
\tilde \rho_{l,t_m}&=&
      (\tilde B_{l,t_m} +\tilde B_{l}\tilde B_{m})
      \tilde \Psi\tilde \Psi^*-\tilde B^*_{l}\tilde \Psi^* \tilde B_m
           \tilde \Psi \nonumber \\ &-&
      (\tilde B^*_{l,t_m} -\tilde B^*_{l}\tilde B^*_{m})
      \tilde \Psi\tilde \Psi^*-\tilde B_{l}\tilde \Psi \tilde B^*_m
           \tilde \Psi^*
.\end {eqnarray}
Invoking now the compatibility condition (2.6) for the gauge transformed
system as
\[
  \tilde B_{l,t_m} +\tilde B_{l}\tilde B_{m}=
  \tilde B_{m,t_l} +\tilde B_{m}\tilde B_{l}\]
and its conjugate
 \[
  \tilde B^*_{l,t_m} -\tilde B^*_{l}\tilde B_{m}^*=
  \tilde B^*_{m,t_l} -\tilde B^*_{m}\tilde B^*_{l}\]
 we can flip the order of differentiation in (3.4):
     \begin {eqnarray}
   \partial_{t_k}
\tilde \rho_{l,t_m}&=&
      (\tilde B_{m,t_l} +\tilde B_{m}\tilde B_{l})
      \tilde \Psi\tilde\Psi^*-\tilde B^*_{m}\tilde\Psi^* \tilde B_l
           \tilde\Psi \nonumber \\ &-&
      (\tilde B^*_{m,t_l} -\tilde B^*_{m}\tilde B^*_{l})
      \tilde \Psi\tilde \Psi^*-\tilde B_{m}\tilde \Psi \tilde B^*_l
           \tilde\Psi^*  \nonumber \\ &=&
( \tilde \Psi\tilde \Psi^*)_{t_m,t_l} \nonumber\\&=&
   \partial_{t_k} \tilde \rho_{m,t_l}
\end {eqnarray}
 yielding finally the required relation between conserved
densities as
         \begin {equation}
\tilde \rho_{l,t_m}=  \tilde \rho_{m,t_l}
\end {equation}
for all constraints (3.2).
Now we can make a consistent choice of gauge function $\theta$ by setting
          \begin {equation}
\theta_{t_l}=  \tilde \rho_{l}
,\end {equation}
 which clearly satisfies the compatility condition
    $\theta_{t_it_j}=  \theta_{t_jt_i} $
   due to the property (3.6).
   The explicit forms of  $\tilde \rho_l$
   can be easily calculated using the known expressions for the
   conserved densities  [3].
   Significantly the choice (3.7) remains valid
   for  the hierarchies of constraint (3.2)
   with any $k$ and using  (3.1-2)
   may be given as
 \begin{equation}
 \theta_{t_l}
= \partial_{t_k}^{-1} (\tilde \Psi\tilde\Psi^*)_{t_l}
\end {equation}
for constraints $k=1,2,3,\cdots$. For example for $k=1$, i.e. for
the constraint (2.9), from (3.7) and (3.8) we get
           \begin {equation}
 {\theta}_x   =  \tilde \Psi \tilde \Psi^* = \tilde \rho_1\ , \ \
{\theta}_y   =  \partial_x^{-1}(\tilde \Psi \tilde \Psi^*)_y =
 \tilde \rho_2\ , \ \
{\theta}_t =  \partial_x^{-1}(\tilde \Psi \tilde \Psi^*)_t =
 \tilde \rho_3
\end {equation}
etc., where $\tilde \rho_l$ for this system are  given by
  \begin {eqnarray}
{\tilde \rho}_1  &=&  \tilde \Psi \tilde \Psi^*
 ,  \\
{\tilde \rho}_2  &=&    (\tilde \Psi_x\tilde \Psi^*-\tilde \Psi\tilde \Psi^*_x)
+
2 (\tilde \Psi\tilde \Psi^*)^2,
    \\ {\tilde \rho}_3 &=& (\tilde \Psi_{xx} \tilde \Psi^*+
   \tilde \Psi\tilde \Psi^*_{xx}  -\tilde \Psi_x \tilde \Psi^*_{x}
+
3 (\tilde \Psi\tilde \Psi^*)^2
\nonumber \\ &+& 3      ( \tilde \Psi\tilde \Psi^* )
  (\tilde \Psi_x\tilde \Psi^*-\tilde \Psi\tilde \Psi^*_x)
+  3
 (\tilde \Psi\tilde \Psi^*)^3
\end{eqnarray}
etc., substantiating thus the choice (2.15-17).

We  start now considering other constraints.
   For  (3.2) with $k=2$ , i.e
for the constraint
     \begin {equation}
 u_y=   ( \tilde \Psi \tilde \Psi^*)_x
\end {equation}
using again  (3.7-8) and the conserved densities listed in ref.3
we get
  \begin {eqnarray}
{\theta}_x   =  \partial_y^{-1}(\tilde \Psi \tilde \Psi^*)_x =
{\tilde \rho}_1  &=& u
 ,  \\
{\theta}_y   =  (\tilde \Psi \tilde \Psi^*) =
{\tilde \rho}_2  &=&    \tilde \Psi\tilde \Psi^*
,    \\
{\theta}_t=  \partial_y^{-1}(\tilde \Psi \tilde \Psi^*)_t =
{\tilde \rho}_3 &=&  \frac{1}{4} u_{xx}+   \frac{1}{4} u^2+  \frac{3}{4}
 (\tilde \Psi_{x} \tilde \Psi^*-
   \tilde \Psi\tilde \Psi^*_{x})  +  \frac{3}{2}u
     \tilde \Psi \tilde \Psi^*
\end{eqnarray}
etc., where $u$ may be replaced everywhere by  (3.13).
 For obtaining the gauge transformed hierarchy of equations, one should
  insert
the gauge choice (3.14-16) into the transformed set (2.11-14) and
impose the constraint (3.13) under consideration. This would yield finally
the system
    \begin {equation}
\tilde  \Psi_{y}= \tilde  \Psi_{xx}+2 u
 \tilde  \Psi +\left[(
 u^2 +{u}_x  -\tilde  \Psi\tilde  \Psi^*    )\tilde  \Psi
     +2u
\tilde  \Psi_x \right]
\end {equation}
along with a similar equation for $\Psi^*$  and
complemented naturally by $ u_y=   (\Psi \Psi^*)_x
.  $
  We are not presenting here the
higher  evolution equations of this hierarchy due to their lengthy forms,
and due to their otherwise straightforward derivation following the
path  indicated above. Note that the terms appearing within the
square brackets
  in (3.17)
are the additional higher-order nonlinear terms to the Yaijima-Oikawa
system and consequently, the gauge generated system in this case
represents a higher-order Yaijima-Oikawa hierarchy.

 The next higher constraint  given by (3.2) with $k=3$ takes the form
            \begin {equation}
  u_t =   (\tilde \Psi\tilde\Psi^*)_x
\ ,     \end{equation}
or using the KP equation  as
           \begin {equation}
  u_y =   \frac{2}{3} v_x
\ , \qquad v_y= -   \frac{1}{2} u_{xxx}  -6uu_x +2
  (\tilde \Psi\tilde\Psi^*)_x    . \end{equation}
Taking into account the conserved densities for this case [3]
we can fix the gauge as
            \begin {eqnarray}
 {\theta}_x   =   \tilde \rho_1 &=&  u \ , \\
{\theta}_y   =
 \tilde \rho_2 &=& v\ , \\
{\theta}_t =
 \tilde \rho_3 &=&       \tilde \Psi \tilde \Psi^*
,\end {eqnarray}
which give from the gauge transformed system (2.11-14)
the  equations
\begin{equation}
\tilde  \Psi_{y}= \tilde  \Psi_{xx}+2 u
 \tilde  \Psi +\left[(
 -v +u^2  +u_{x} )\tilde  \Psi
     +2u
\tilde  \Psi_x       \right]
,\end{equation}
\begin{equation}
\tilde  \Psi_{y}^*= -\tilde  \Psi_{xx}^* -2 u
 \tilde  \Psi^* +\left[(
 v -u^2  +u_{x} )\tilde  \Psi^*
     +2u
\tilde  \Psi_x^*       \right]
,\end{equation}
together with the constraint equations (3.19)
and  for $t$-evolution :
\begin{eqnarray}
\tilde{\Psi}_t& = &\tilde{\Psi}_{xxx} + 3u\tilde{\Psi}_x +
                \frac{3}{2}u_x \tilde{\Psi}+
             v   \tilde{\Psi} \nonumber \\&+& \left[
                 (-(\tilde \Psi \tilde \Psi^*) +3u^2 +u_{xx}
                 +3u u_{x}+u^3)\tilde \Psi
                  + 3 (u_{x}
                  +u^2)\tilde \Psi_x+3u
                 \tilde \Psi_{xx} \right] \ ,
                  \end{eqnarray}
                   \begin{eqnarray}
\tilde{\Psi}^*_t& = &\tilde {\Psi}^*_{xxx} + 3u\tilde {\Psi}^*_x +
                  \frac{3}{2}u_x \tilde {\Psi}^*-
            v      \tilde {\Psi^*}
                  \nonumber \\&-&          \left[
                 (-\tilde \Psi \tilde \Psi^* +3u^2+u_{xx}
                 -3u u_{x}+u^3)\tilde \Psi^*
                 -3(u_{x} -u^2)
                  \tilde \Psi_x^*-3 u
                 \tilde \Psi_{xx}^*  \right]
\end{eqnarray}
along with the  $t$-evolution
 equations given by the constraint  (3.18) and
            \begin {equation}
  v_t =  \frac{3}{2} (\tilde \Psi_{xx}\tilde\Psi^*-
   \tilde \Psi\tilde\Psi^*_{xx})+\left[3(u_x  (\tilde \Psi\tilde\Psi^*) +u
    (\tilde \Psi\tilde\Psi^*)_x)       \right]
\ ,                                             \end{equation}
and similarly for higher evolutions.  We should note again that the
terms marked in  square brackets in  equations  (3.23-27)
are the extensions  to the Melnikov hierarchy
, which becomes clear after comparison
with the known result [3] and therefore the integrable  system
generated here can be considered as a new  extended Melnikov hierarchy.

\section {
Multicomponent generalisations
}
\setcounter{equation}{0}

 It has been shown that the constraints in KP hierarchy
 can be extended without much effort to vector eigenfunctions [3].
 This may be achieved by considering simply $\Psi,\Psi^*$ to be
 multicomponent objects  $\Psi^i,\Psi^{i*}, \  i=1,2,\cdots m$
  and replacing (3.2) by
 \begin {equation}
  \partial_{t_k}u =  \partial_{x} (\sum_{i=1}^m
   \Psi^i\Psi^{i*})
\ , \qquad\ k=1,2,3,\cdots.\end {equation}
     producing the series of constraits (1.1). Under such
      extension  the vector generalisations of the
      original  constrained systems ,e.g. vector AKNS, vector
       Yaijima-Oikawa, vector Melnikov etc. hierarchies are obtained
        [3].
       That is for the simplest constraint
$ u =   (\sum_{i=1}^m
  \Psi^i \Psi^{i*})
,$
  one gets in particular the vector NLS-type (AKNS) equation
   \begin{equation}
\vec  \Psi_{y}= \vec  \Psi_{xx}+2   <\vec  \Psi \cdot\vec  \Psi^\dagger>
 \vec \Psi
.\end{equation}
 with the  notation $  <\vec  \Psi \cdot\vec  \Psi^\dagger>
=\sum_{i=1}^m
  \Psi^i \Psi^{i*}   $  for the scalar-product.

 With the aim of applying the gauge generation procedure developed
 in previous sections to this multicomponent case , we notice
 first that the gauge freedom involved in  constraints (4.1) has become
 much wider.Each component of the eigenfunctions can be changed
 now independently as
   $ \tilde \Psi^i=e^{-\theta^i} \Psi^{i}
    ,\ \Psi^{i*}= \Psi^{i*}
 e^{\theta^i}, \quad i=1,2\cdots m , \ $
   with $m$ number of arbitrary functions $\theta^i$. We show
    below that such a
freedom induces interesting multicomponent generalisations of the
higher-order extended hierarchies of   the constrained equations already
obtained  in the single component case in sec. 2-3.
One can see that in multicomponent case the gauge transformed equations
for each component can be given again by the same  (2.11-14)
 by replacing    only
      \begin{equation}
\tilde  \Psi \rightarrow \tilde  \Psi^i , \ \tilde  \Psi^{*}
 \rightarrow \tilde  \Psi^{i*} , \quad \theta \rightarrow \theta^i
\end{equation}
    in these equations. For the corresponding choice of the gauge
    functions $\theta^i$ we may use the expressions
 \begin{equation}
 \theta^i_{t_l}
= \partial_{t_k}^{-1} (\tilde \Psi^i\tilde\Psi^{i*})_{t_l}
\end {equation}
    parallel to (3.8). However we should note here that individual
     $\theta^i$
    can not be expressed  any longer through the conserved densities
    , as has been done  in the single component case. Such functions
now    are  related only to a part of the total densities as
    $ \ \sum_{i=1}^m   \theta^i_{t_l} = \tilde \rho_l \ $. Nevertheless
    due to obvious structure of (4.4)  the compatibility
    $ \theta^i_{t_l,t_m}  = \theta^i_{t_m,t_l}  $ holds and more
     than that
    no other $ \theta^j_{t_m} $'s appear in the expressions
    of $ \theta^i_{t_l} $ , which  enables us to
    choose (4.4) without difficulty . This important property of nonmixing
    between $ \theta^i_{t_l}$'s  may be explained by the fact that
    the original (unconstrained) system (2.1) for $ \Psi^i  $
    was afterall a linear system  with coefficients expressed in
    $u_i, u_{i,t_l}$, which remain unaffected by the gauge transformation.
    Therefore due to this excellent property all the formulas derived
    for the single component case in sec.2-3 continue to be  valid
    also in the multicomponent case after the replacement (4.3).
    As a result for example, for the simplest constraint
         \begin {equation}
  u =   (\sum_{i=1}^m
  \tilde \Psi^i\tilde\Psi^{i*})
 \end {equation}
we may choose
       \begin {eqnarray}
{\theta}_x^i  &=&  \tilde \Psi^i \tilde \Psi^{i*}
 ,  \\
{\theta}_y^i  &=&    (\tilde \Psi_x^i\tilde \Psi^{i*}
-\tilde \Psi^i\tilde \Psi^{i*}_x)
+
2 (\tilde \Psi^i\tilde \Psi^{i*})^2,
    \\ {\theta}_t^i &=&
    \tilde \Psi_{xx}^i \tilde \Psi^{i*}+
   \tilde \Psi^i\tilde \Psi^{i*}_{xx}  -\tilde \Psi_x^i \tilde \Psi^{i*}_{x}
+
3(\sum_{j=1}^m \tilde \Psi^j\tilde \Psi^{j*})
 (\tilde \Psi^i\tilde \Psi^{i*})
\nonumber \\ &+& 3      ( \tilde \Psi^i\tilde \Psi^{i*} )
  (\tilde \Psi_x^i\tilde \Psi^{i*}-\tilde \Psi^i\tilde \Psi^{i*}_x)
+  3
 (\tilde \Psi^i\tilde \Psi^{i*})^3
,\end{eqnarray}
using (2.15-17) ( $\theta^i_x$ is nolonger expressible through
$u$!), yielding a multicomponent generalisation of Kundu-Eckhaus
(MKE) equation
  \begin{equation}
\tilde  \Psi_{y}^i= \tilde  \Psi_{xx}^i+   2 \sum_{j=1}^m
(\tilde  \Psi^j\tilde  \Psi^{j*})  \tilde  \Psi^i
+ \left[2(\tilde  \Psi^i\tilde  \Psi^{i*})_x -(\tilde  \Psi^i\tilde
\Psi^{i*})^2
\right] \tilde  \Psi^i
,\end{equation}
an analogous one for $\Psi^{i*}$ and similarly  for the higher evolutions.
It is significant to observe that   in  the MKE equation obtained above
the original global $SU(m)$ symmetry has been broken down to just $U(1)$
 due to the explicit presence of anisotropic extensions
and as a consequence,  such multicomponent equations can not be written
in a vector form. It is easy to  notice that under the $SU(m)$
 transformation
\begin{equation}
\vec {\tilde \Psi}'=\vec {\tilde \Psi} U, \  \ \vec{\tilde \Psi^\dagger}'
= U^\dagger\vec{ \tilde \Psi^\dagger}
, \qquad UU^\dagger=I     \end{equation}
in (4.9), only the part outside the square brackets (i.e. the original vector
 AKNS equation) remains invariant.

 For other constraints ($k=2,3,\cdots$)
   following the above steps ,  one can also get similar multicomponent
  generalisations of higher-order Yijima-Oikawa and Melnikov hierarchies
  exhibiting the same type of symmetry breaking
. It is evident  that in such cases one has different possibilities
of anisotropic
 reductions by choosing any number of $\theta^l, \ l\in[1,m]$ to be
trivial and thus extending  only those equations with
nontrivial $\theta^m, \ m\neq
l$  to higher-order, while leaving the rest   to follow
 the same original
equations ,e.g. AKNS,Yijima-Oikawa , Melnikov equations etc.
However there can be  a qualitatively different possibility , when
all $\theta^i
$ are given by the same function : $\theta^i=\theta , \ i\in[1,m]$.
For this  isotropic gauge transformation we may choose
     \begin{equation}
 \theta_{t_l}
= \partial_{t_k}^{-1}
<\vec{\tilde \Psi}\cdot\vec{\tilde \Psi}^{\dagger}>
=\tilde \rho_{t_l}
,    \end{equation}
expressed through the scalar-product of  $\vec{\tilde \Psi}
$. It is evidently different from (4.6-8) and expressible again
through the conserved densities of the multicomponent constrained
 system  as well as through  $u$. As a cosequence the global
  symmetry is
 restored again
  in the gauge extended system and  the constraint
 (4.5)
  yields now a genuine vector Kundu-Eckhaus (VKE) equation
in the form  \begin{eqnarray}
\vec {\tilde  \Psi}_{y}&=&\vec {\tilde  \Psi}_{xx}+(2
<\vec{\tilde \Psi}\cdot\vec{\tilde \Psi}^{\dagger}>
- <\vec{\tilde \Psi}\cdot\vec{\tilde \Psi}^{\dagger}>^2
)\vec {\tilde  \Psi}\nonumber \\
& +&2 <\vec{\tilde \Psi}\cdot\vec{\tilde \Psi}^{\dagger}>
 \vec{\tilde \Psi}_x+
  2 <\vec{\tilde \Psi}\cdot\vec{\tilde \Psi}^{\dagger}_x>
 \vec{\tilde \Psi}
,\end{eqnarray}
 obviously having higher symmetry than
the  MKE equation (4.9). It should be noticed also that the
 VKE equation , which is the vector generalisation of (2.18) exhibits
slightly different structure compared to (2.18) ,
 since the last two terms in (4.12) can
not be combined
together due to their vector form. Such vector generalisations
preserving the global symmetry also go through for higher constraints
yielding similarly vector extensions of higer-order Yijima-Oikawa and
Melnikov hierarchies.
\section{Gauge generation from constrained modified KP hierarchy
}
\setcounter{equation}{0}
                     Recently [7,8] it has been shown  that
    for the modified KP hierarchy compatible with the linear system

\begin{equation}
  \Psi_{y}=   \Psi_{xx}-2 v
  \Psi_x
\end{equation}
and its conjugate
\begin{equation}
 \Psi_{y}^*= - \Psi_{xx}^* -2 v
  \Psi^*
\end{equation}
along with the higher evolution equations , one may impose the constraint
$  v= \Psi  \Psi^* $ in (5.1-2) resulting the Chen-Lee-Liu  equation
    \begin{equation}
  \Psi_{y}=   \Psi_{xx}-2 (\Psi  \Psi^* )
  \Psi_x
\end{equation}
    and its conjugate and their hierarchy.
Our aim is to demonstrate
 that the gauge generation formulated for the constrained
KP system is  applicable also to this case.

it may be shown that for the gauge transformation
   $ \tilde \Psi=e^{-\alpha \theta} \Psi ,\ \tilde \Psi^{*}= \Psi^{*}
 e^{\alpha\theta} $
  , where $\alpha$ is a constant, the gauge function $\theta$ in this case
  may be chosen consistently as
       \begin {eqnarray}
{\theta}_x  &=&   \tilde \Psi \tilde \Psi^*
 ,  \\
{\theta}_y  &=&    (\tilde \Psi_x\tilde \Psi^*-\tilde \Psi\tilde \Psi^*_x)
+
(2\alpha-1) (\tilde \Psi\tilde \Psi^*)^2,
    \\ {\theta}_t &=&
    \tilde \Psi_{xx} \tilde \Psi^*+
   \tilde \Psi\tilde \Psi^*_{xx}  -\tilde \Psi_x \tilde \Psi^*_{x}
+
3(\alpha-1)^2 (\tilde \Psi\tilde \Psi^*)^2
\nonumber \\ &+& 3 (1+\alpha)     ( \tilde \Psi\tilde \Psi^* )
  (\tilde \Psi_x\tilde \Psi^*-\tilde \Psi\tilde \Psi^*_x)
+  3
 (\tilde \Psi\tilde \Psi^*)^3
.\end{eqnarray}
 Under such gauge transformation the constraint remains invariant:
        \begin{equation}
  v= (\tilde\Psi  \tilde\Psi^* )
,\end{equation}
 while the system (5.1-2) changes to
\begin{equation}
\tilde  \Psi_{y}= \tilde  \Psi_{xx}-2 (\Psi  \Psi^* )
 \tilde  \Psi_x + \alpha(
 -{\theta}_y + \alpha{\theta}_x^2  +{\theta}_{xx}  -2\theta_x
 (\tilde \Psi \tilde \Psi^* ) )\tilde  \Psi
     +2\alpha{\theta}_x
\tilde  \Psi_x
,\end{equation}
 along with  its conjugate, which  using (5.4-6) finally
 yields the equation
 \begin{equation}
\tilde  \Psi_{y}= \tilde  \Psi_{xx}+2 (\alpha-1)
(\tilde  \Psi\tilde  \Psi^*) \Psi_x
+ 2\alpha \tilde  \Psi^2\tilde  \Psi^*_x -\alpha(1+\alpha)
(\tilde  \Psi\tilde  \Psi^*)^2
 \tilde  \Psi
.\end{equation}
Similarly the conjugate and higher evolution equations are also
obtained.
Obviously for $\alpha=0$, when the effect of gauge transformation
is removed one gets back the Chen-Lee-Liu equation (5.3). However in
general  (5.9) is a higher-order nonlinear extension and
more interestingly for $\alpha=-1 $ it reduces to the standard Kaup-Newell
 derivative NLS equation [9]
 \begin{equation}
\tilde  \Psi_{y}= \tilde  \Psi_{xx}
-2\left((\tilde  \Psi\tilde  \Psi^*) \Psi\right)_x
,\end{equation}
while for $\alpha=+1$ to the Gerdjikov-Ivanov [10] equation
    \begin{equation}
\tilde  \Psi_{y}= \tilde  \Psi_{xx}
+ 2 \tilde  \Psi^2\tilde  \Psi^*_x -2
(\tilde  \Psi\tilde  \Psi^*)^2
 \tilde  \Psi
.\end{equation}
     Thus the constrained modified KP hierarchy is equivalent to
     the hierarchy of generalised derivative NLS
      equation (5.9), which reduces to different
     other known hierarchies depending on the gauge choice.
\section{ Concluding remarks}
\setcounter{equation}{0}
The known constrained KP  or modified KP hierarchy does not exhaust
the gauge freedom allowed by the series of constraints in these systems.
Such gauge freedom can be exploited effectively to generate new
integrable  higher-order
nonlinear extensions to the known hierarchies. Thus
the Kundu-Eckhaus hierarchy,
 higher-order Yijima-Oikawa and Melnikov hierarchy
as well as the generalised derivative NLS  equation leading to
the standard
derivative NLS , the Gerdjikov-Ivanov and the Chen-Lee-Liu
 equations, etc. are constructed.
Multicomponent extension  yields two qualitatively  different kinds of
generalisation  of the above hierarchies. In the isotropic vector
 generalisation the global symmetry of the original system is preserved,
 while in other anisotropic multicomponent extensions the original
 symmetry is heavily broken down. The preservation of integrability
in the extended models inspite of   the breaking down of symmetries
observed here is a significant fact,
since  commonly one encounters quite opposite  picture  [13].

 \section*{ Acknowledgement}
One of the authors (AK) likes to express
 his thanks to the Alexander von Humboldt Foundation
for  its fellowship award.


\section*{References}
\begin{enumerate}
\item Y.Cheng and Y.S.Li, Phys. Lett. A, 157, 22 (1991).
\item B.G.Konopelchenko, J.Sidorenko and W.Strampp,
      Phys. Lett. A, 157, 17 (1991).
\item      J.Sidorenko and W.Strampp, J. Math. Phys., 34, 1429 (1993).
\item A.Kundu, J. Math. Phys., 25,  3433 (1984).
\item F.Calogero, Inverse Prob., 3, 229 (1987)
\item Li shen in 'Symmetries and Singularity Structures' (Sringer Publ.
, 1990,ed. M.Lakshmanan), p.27
\item Yi Cheng and Yi-shen Li , J. Phys. A, 25, 419 (1992).
\item W.Oevel and W.Schief,
      Squared eigenfunctions of the (modified) KP hierarchy
      and scattering problems of Loewner type.
      Applied Mathematics Preprint AM 93/21,
      University of New South Wales, Australia (1993).

\item D.J.Kaup and A.C.Newell, J. Math. Phys., 19, 789, (1978)
\item V.S.Gerdjikov and M.I.Ivanov, Bulg. J. Phys., 10, 13 (1983).
 \item   H. H. Chen ,  Y. C. Lee and C.S. Liu, Phys. Scripta, 20 ,490 (1979).
\item Y.Ohta, J.Satsuma, D.Takahashi and T.Tokihiro, Prog. Theor. Phys.
      Suppl., 94, 219 (1988).
\item V.G.Makhankov, Phys. Rep., 35, 1 (1978).

\end{enumerate}
\end{document}